\documentclass[lettersize,journal]{IEEEtran}

\usepackage{amsmath}
\usepackage[]{graphicx}
\usepackage[]{tabularx}
\DeclareGraphicsExtensions{.pdf,.jpeg,.png}
\usepackage{subfig}
\usepackage{xspace}
\usepackage{xcolor}
\usepackage{paralist}
\usepackage{color}
\usepackage{listings}
\usepackage{listings-golang}
\usepackage{float}
\usepackage{multirow}
\usepackage{url}

\newcommand{\fig}[1]{Fig.~\ref{fig:#1}}
\newcommand{\tab}[1]{Table~\ref{tab:#1}}
\newcommand{\s}[1]{Section~\ref{sec:#1}}

\newcommand{\fakeparagraph}[1]{\vspace{2mm}\noindent\textbf{\textit{#1.}}}

\newif\ifhidenotes
\hidenotesfalse

\ifhidenotes
\newcommand{\note}[1]{}
\newcommand{\noteam}[1]{}
\newcommand{\noteldm}[1]{}
\else
\newcommand{\note}[1]{\footnote{\color{purple}{Note: #1}}}
\newcommand{\noteam}[1]{\footnote{\color{blue}{Ale: #1}}}
\newcommand{\noteldm}[1]{\footnote{\color{red}{Luca: #1}}}
\fi

\newcommand{\sysname}{Histrio\xspace}

\begin{document}

\title{\sysname: a Serverless Actor System}

\author{Giorgio Natale Buttiglieri, Luca De Martini, Alessandro Margara%
  \IEEEcompsocitemizethanks{\IEEEcompsocthanksitem Dipartimento di
    Elettronica, Informazione e Bioingegneria%
    \protect\\Politecnico di Milano%
    \protect\\mail: \{luca.demartini, alessandro.margara\}@polimi.it,%
    \protect\\\{giorgionatale.buttiglieri\}@mail.polimi.it}%
  \thanks{Manuscript received October 25, 2024.}}

  

\maketitle

\begin{abstract}
  In recent years, the serverless paradigm has been widely adopted to develop
cloud applications, as it enables building scalable solutions while delegating
operational concerns such as infrastructure management and resource
provisioning to the serverless provider.
Despite bringing undisputed advantages, the serverless model requires a change
in programming paradigm that may add complexity in software development.
In particular, in the Function-as-a-Service (FaaS) paradigm, functions are
inherently stateless: they perform actions without retaining any state.
As a consequence, developers carry the burden of directly interacting with
external storage services and handling concurrency and state consistency
across function invocations.  This results in less time spent on solving the
actual business problems they face.

Moving from these premises, this paper proposes \sysname, a programming model
and execution environment that simplifies the development of complex stateful
applications in the FaaS paradigm.
\sysname grounds on the actor programming model, and lifts concerns such as
state management, database interaction, and concurrency handling from
developers.
It enriches the actor model with features that simplify and optimize the
interaction with external storage.
It guarantees exactly-once-processing consistency, meaning that the
application always behaves as if any interaction with external clients was
processed once and only once, masking failures.

\sysname has been compared with a classical FaaS implementation to evaluate
both the development time saved due to the guarantees the system offers and
the applicability of \sysname in typical applications.
In the evaluated scenarios, \sysname simplified the implementation by
significantly removing the amount of code needed to handle operational
concerns.  It proves to be scalable and it provides configuration mechanisms
to trade performance and execution costs.
\end{abstract}


\section{Introduction}
\label{sec:intro}

In recent times, there has been an increasing interest in the serverless
computing model.
Indeed, this model brings advantages in terms of development experience,
management, and pricing. When developing for serverless, developers only need
to write the application logic, without worrying about its deployment, which
is handled by the service provider, with automatic scaling and pay-as-you-go
price models.
In the context of serverless computing, the function-as-a-service paradigm
(FaaS) implements applications as a composition of functions, delegating their
scheduling and execution to the serverless environment.

Despite bringing undisputed advantages, the serverless model requires a change
in programming paradigm that may add complexity in software development.
In particular, in the FaaS paradigm, functions are inherently stateless: they
perform actions without retaining any state. This is very different with
respect to the traditional imperative programming style and requires
developers to adapt to this new environment: instead of directly manipulating
state through variables, developers must now explicitly encode interactions
with external storage systems --- typically a managed database --- to persist
the state of their applications.

So far, little effort has been put to provide abstractions and paradigms that
aid developers in modeling and implementing solutions that deal with the
inconveniences of serverless paradigms.
Moving from these premises, this work proposes \sysname, a programming model
and execution environment that simplifies the development of complex stateful
applications in the FaaS paradigm.

\sysname grounds on the already established actor
model~\cite{actorModelDefinition}.  It lifts concerns such as state
management, database interaction, and concurrency handling from developers.
Developers model the domain problem using actors, which are independent
stateful entities that interact with each other only through the exchange of
messages.
\sysname automatically and implicitly generates the code to execute the
application in a serverless environment. It runs the code that implements the
behavior of actors with the FaaS model, it manages the communication between
actors, it persists the state of actors within an external storage service,
automating and hiding state access.
\sysname enriches the actor model with query-like features that facilitate and
optimize the most frequent state access patterns without losing the simplicity
of its interface.
Overall, these features enable developers to focus on the business problem
they face and forget about low-level concerns.

Furthermore, \sysname also addresses another of the common problems of
serverless; as functions may be randomly terminated and resumed by the service
provider,
there are no guarantees that an application handles requests from external
clients as expected.
\sysname overcomes this problem by implementing a failure handling mechanism
that guarantees exactly-once-processing consistency system-wide, meaning that
the application is guaranteed to behave as if any interaction from external
clients was processed once and only once, even in the case of failures.

We implemented the \sysname environment using the AWS Lambda as a FaaS
environment and DynamoDB as a storage service to persist state.
The paper discusses in detail the programming model of \sysname and its
implementation as a serverless platform.
It describes the protocols that govern the interaction with the storage
service and provide exactly-once-processing guarantees.
To evaluate \sysname, we implemented two real-world applications using
\sysname and using a baseline approach that relies on classic serverless
abstractions.  We show that \sysname significantly simplifies the development
of applications by rising the level of abstractions for developers.  It
introduces a latency overhead that is configurable based on the specific needs
of applications: developers can trade response time for reduced costs, always
keeping the budget for executing their applications under control.

The paper is structured as follows: \s{background} introduces background
concepts and motivations to the work.  \s{api} explores the programming
paradigm offered by \sysname.  \s{model} presents an overview of the system
and \s{impl} shows implementation details of its components.  \s{eval}
contains an evaluation in terms of performance and code complexity with a
discussion of the results.  \s{related} discusses related work, and
\s{conclusions} provides conclusive remarks.

\section{Background and Motivations}
\label{sec:background}

This section explores the technological landscape around \sysname and the
actor model it is inspired by.  In the end, it summarizes the motivations that
lead us to develop \sysname.

\subsection{Serverless}

Serverless is a cloud-native development model that allows developers to build
and run applications without having to manage
servers~\cite{serverlessDefinition}. This model encompasses different services
that can be accessed without directly worrying about provisioning,
maintenance, or scaling. These services usually adhere to a pay-as-you-go
paradigm, so the upfront investment is lower compared to an on-premise
infrastructure.

\fakeparagraph{FaaS}
Function-as-a-Service, or FaaS, is a kind of cloud computing service that
allows developers to build, compute, run, and manage application packages as
functions without having to maintain their own
infrastructure~\cite{faasDefinition, faasImplications}.

\sysname builds on AWS Lambda~\cite{awsLambda} as a provider for the execution
of functions.  AWS Lambda is one of the most common FaaS solutions, supporting
packages written in many languages.
When a function is invoked, the Lambda service forwards the request to a
suitable function instance, which is a container running the package uploaded
for that function. If no function instance suitable for the processing of the
request is available, Lambda allocates a new one to handle the request.
A function instance can handle only one request at a time, hence Lambda scales
its processing capabilities by creating new instances upon receiving a growing
request traffic.
The cost of using Lambda depends on the number of invocations for each
function, the RAM/cores allocated for containers and the duration of the
invocations.

While the current implementation of \sysname builds on AWS Lambda, all main
public cloud vendors offer equivalent solutions that expose the same
programming model.  Examples include Microsoft Azure
Functions~\cite{azureFunctions}, Google Cloud Functions~\cite{cloudFunctions},
and Cloudflare Workers~\cite{cloudflareWorkers}

\fakeparagraph{Serverless databases}
A serverless database is managed by a third-party cloud
provider~\cite{serverlessDatabaseDefinition}.
\sysname adopts DynamoDB~\cite{awsDynamoDB} as its storage layer.  DynamoDB is
a NoSQL serverless database offered by AWS, frequently used in combination
with AWS Lambda functions.
It was chosen due to its low latency~\cite{dynamodbLatency} and the
consistency guarantees it provides, which match the requirements of \sysname
without introducing a significant performance overhead.

In a nutshell, DynamoDB is a distributed storage service consisting of
multiple \emph{storage nodes}.  It offers a key-value model, where data is
partitioned across storage nodes using a \emph{partition key} and sorted
within each node according to a \emph{sort key}.
Routers dispatch client requests to the storage nodes.
Each data item has a leader storage node that handles all write operations,
and a configurable number of replica nodes that are updated after a write has
been performed on the leader.
Read operations are eventually consistent by default, but can be configured to
be strongly consistent on a per-request basis.
One key feature of DynamoDB is that it offers the possibility to perform
conditional update operations. This feature can be used akin to
compare-and-swap operations to implement external synchronization patterns to
be used by applications.
Lastly, the write transactions allow bundling multiple write operations in a
single request, which guarantees that either all or none of the changes are
atomically applied to the database.
\sysname uses conditional updates and write transactions as the main tools to
guarantee exactly-once processing semantics.

\subsection{Actor model}

The actor model is a mathematical model of concurrent computation that treats
an \emph{actor} as the basic building block of concurrent
computation~\cite{actorModelDefinition}.
An actor is an entity that encapsulates some private state and interacts with
other actors by sending and receiving messages.

When processing a message, an actor can:
\begin{inparaenum}[(i)]
\item modify the private state owned by the actor;
\item spawn some new actors;
\item send messages to other actors.
\end{inparaenum}
Each actor processes one message at a time, ensuring that no race conditions
are possible on its state.
Concurrency emerges by letting multiple actors execute simultaneously. This
leads to scalable concurrent systems, where the unit of concurrent computation
is the actor itself. When needed, new actors can be spawned allowing the
system to withstand a growing workload

\subsection{Motivations and executive summary}

Although FaaS and serverless databases help developers to build applications
without the burden of direct resource management, they lack some guarantees
frequently needed when developing distributed applications, and in particular
Web applications.

In this context, it is common that multiple serverless function instances need
to access and modify a shared state concurrently.
As for AWS Lambda, the only way to control concurrency is to limit the maximum
number of instances a function can have at any time. Other than that, no other
mechanism is provided to handle coordination between instances of a
function. Lambda does not offer any synchronization API, so developers need to
address this concern by themselves, usually relying on features offered by the
data store they use.

Fault tolerance is another concern that is not directly addressed by typical
FaaS environments. Appropriate fault tolerance guarantees are useful because,
if not present, developers need to write code to ensure correctness despite
failures.
AWS Lambda offers a retry mechanism for faults and application errors. This
approach however does not ensure the complete execution of the function code,
and it assumes the function to be idempotent, which is another concern
developers need to ensure.

In summary, to provide its advantages, the serverless shifts many
responsibilities on the developer.
\sysname aims to solve these problems by:

\begin{enumerate}
\item grounding on the actor model as a solid foundation to build concurrent
  applications;
\item using source-to-source compilation to abstract the management of state
  and hide the interaction with external storage services;
\item providing end-to-end exactly-once semantics for message processing,
  ensuring that, even in the presence of failures, the application behaves as
  if each message was processed and produced any side effects involving the
  state of actors once and only once.
\end{enumerate}

\section{Programming Interface}
\label{sec:api}

This section describes the programming interface of \sysname. We outline the
main features offered by \sysname, starting from the definition of an actor,
its state, and its behavior.

\subsection{Actors: definition and communication}

\sysname is implemented in the Go programming language, a common choice to
build distributed applications\footnote{Current implementation available at
\url{https://github.com/deib-polimi/histrio}}. In \sysname an actor is any
object that implements the following interface:

\begin{lstlisting}
type Actor interface {
  ReceiveMessage(message Message) error
  GetId() ActorId
  SetId(ActorId)
}
\end{lstlisting}

The \texttt{ReceiveMessage} method defines the behavior of an actor when it
receives a message. \texttt{ActorId} is the identifier of the actor, and it is
composed of three parts:

\begin{enumerate}
\item \texttt{PartitionName}: partitions are logical groups in which actors
  are organized. For example, when modeling a travel booking application, the
  partition for the 'TravelAgency' actors might be the city in which they
  operate. Users need to explicitly define the name of the partition to which
  the actor belongs.
\item \texttt{ShardId}: generated part of the identifier that represents the
  physical shard the actor is assigned to.
\item \texttt{InstanceId}: part of the identifier of the actor chosen by the
  user that needs to be unique inside the same partition.
\end{enumerate}

The code snippet below exemplifies the definition of an actor type
\texttt{TravelAgency}.
The first three lines define the attributes (state) of \texttt{TravelAgency}
actors. Since the state needs to be persisted, the only restriction on the
data that constitutes the state of the actor is that it needs to be
serializable.
In the example, the state of \texttt{TravelAgency} consists of two attributes:
an identifier \texttt{Id}, which will be generated by the \emph{Spawn} method,
and an \texttt{Address} as an example of object data.

\begin{lstlisting}
type TravelAgency struct {
  Id      ActorId
  Address string
}

func (ta *TravelAgency) GetId() ActorId {
  return ta.Id
}

func (ta *TravelAgency) SetId(actorId ActorId) {
  ta.Id = actorId
}
...
\end{lstlisting}

The following code snippet completes the definition of the
\texttt{TravelAgency} actor by introducing a new type of messages
(\texttt{AddressUpdateRequest}) and by implementing the
\texttt{ReceiveMessage} method to update the address of an actor in response
to an \texttt{AddressUpdateRequest} messages.

\begin{lstlisting}
type AddressUpdateRequest struct {
  NewAddress string
}

func (ta *TravelAgency) ReceiveMessage(message Message) error {
  if addressUpdateRequest, ok := message.(*AddressUpdateRequest); ok {
    return ta.updateAddress(*addressUpdateRequest)
  } else {
    return errors.New("unable to process message")
  }
}

func (ta *TravelAgency) updateAddress(addressUpdateRequest AddressUpdateRequest) error {
  ta.Address = addressUpdateRequest.NewAddress
  return nil
}
\end{lstlisting}

An important thing to notice is that updating the state does not require
special code to interact with the database. Indeed, relieving the burden of
state management from the developer, is one of the intended goals of \sysname.

Actors interact with each other by exchanging messages. To do so, they use a
\texttt{MessageSender} that has the following interface:

\begin{lstlisting}
Tell(payload Message, receiver ActorId)
TellExternal(payload Message, externalId string)
\end{lstlisting}

The \texttt{Tell} method accepts a message and a receiver and ensures that
\textit{eventually} the message will be placed in the receiver mailbox.
The \texttt{TellExternal} allows actors to communicate with the external world
by placing a message in an output table with an application-chosen identifier.
External components can include an Id in their request and look for the
response with the same Id in the output table.
Note that there are no guarantees about the moment in which the message will
be delivered, and it is not necessarily true that immediately after the
completion of the \texttt{Tell} method the message is inside the receiver
mailbox. This decouples the execution of the actor code and the
message-sending process. \sysname delays the sending of messages to other
actors up to the end of the processing to comply with the
exactly-once-processing consistency model.

\begin{lstlisting}
type TravelBookingRequest struct {
  TravelerId   ActorId
  JourneyId string
}

type TravelBookingReply struct {
  AgencyId       ActorId
  JourneyId       string
  IsJourneyBooked bool
  FailureReason  string
}

// TravelAgency Actor
type TravelAgency struct {
  Id      ActorId
  Address string  
  MessageSender MessageSender
}

func (ta *TravelAgency) processTravelBookingRequest(travelBooking TravelBookingRequest) error {
  ta.MessageSender.Tell(TravelBookingReply{
    AgencyId:        ta.Id,
    JourneyId:       travelBooking.JourneyId,
    IsJourneyBooked: false,
    FailureReason:   "This agency does not have any journey yet",
  }, travelBooking.TravelerId)
  return nil
}
\end{lstlisting}

The code snippet above exemplifies the use of the \texttt{Tell} method. The
\texttt{TravelBookingRequest} and \texttt{TravelBookingReply} are messages
that represent the request from a user (\texttt{Traveler}) to book a journey
and the response to that request, respectively.
As the example shows, an actor can access a \texttt{MessageSender} just by
declaring it as part of its state. The \sysname execution environment takes
care of injecting a valid \texttt{MessageSender} when the actor begins
processing. The content of the message to be sent can be any serializable
data.

An important property of the \texttt{Tell} method is that it ensures that the
message will be delivered exactly once. This means that the receiver actor
does not need to handle possible lost or duplicate messages, allowing the
developers to focus on the domain problem and not on the infrastructure. As a
matter of fact, in the example above, every line of code defines either data
structures of behaviors that belong to the specific application logic, and all
concerns related to exchange of information between actors are delegated to
the \texttt{Tell} method.

\subsection{Queryable collections}

When modeling a problem it is common to organize homogeneous entities in
collections, so that they can be handled and accessed together. As an example,
a travel agency might have different journeys it offers.
One possible solution to this requirement is to include the list of journeys
in the state of the \texttt{TravelAgency} actor.
This idea is consistent with the model discussed so far, but it has some
drawbacks:
\begin{inparaenum}[(i)]
\item The list of journeys might be large. Recall that \sysname automatically
  persists the state of actors within an external storage service. As a
  consequence, including large collections of data within the state of an
  actor forces the system to load them when the actor state is loaded.
\item Queries against the journeys need to be coded explicitly. For example,
  to find all journeys that satisfy a certain property, the developer will
  need to manually check every element of the list.
\end{inparaenum}

\sysname addresses the above problems by introducing the concept of
\texttt{QueryableCollection}, which is a collection of \texttt{QueryableItem}s
that can be queried against some of their attributes.
\texttt{QueryableCollection}s bring two advantages:
\begin{inparaenum}[(i)]
\item they enable inspecting the list of items with a query-like methodology;
\item they delegate query execution to the database, which is optimized for
  this task.
\end{inparaenum}

In \sysname, a \texttt{QueryableItem} is any type that implements the
following interface:

\begin{lstlisting}
type QueryableItem interface {
  GetId() string
  GetQueryableAttributes() map[string]string
}
\end{lstlisting}

The \texttt{GetId} method returns the identifier of the item. Each item needs
to have a unique identifier inside the collection. This identifier is used to
efficiently find the item in the collection given its Id.
The \texttt{GetQueryableAttributes} method returns the attributes of the item
that can be subject to queries. It returns a map that associates the name of
each attribute in the item to the value of that attribute.
A requirement for queryable attributes is to be convertible to string. Other
than that, a \texttt{QueryableItem} can contain any data as long as it is
serializable.

The following code snippet shows an example of a \texttt{QueryableItem} that
represents a journey in the travel agency scenario:

\begin{lstlisting}
type Journey struct {
  Id                string
  Destination       string
  Cost              float64
  AvailableBookings int
}

func (t *Journey) GetId() string {
  return t.Id
}

func (t *Journey) GetQueryableAttributes() map[string]string {
  return map[string]string{
    "Destination": t.Destination,
  }
}
\end{lstlisting}

Collections of \texttt{Journey}s support the following access pattern:

\begin{itemize}
\item Get one journey given its Id;
\item Find the journeys that satisfy some condition on the destination attribute.
\end{itemize}

\fakeparagraph{Get}
The following code snippet complements the previous example adding a
\texttt{QueryableCollection} of \texttt{Journey}s to the \texttt{TravelAgency}
actor and completing the interaction of a booking request from a traveler.

\begin{lstlisting}
type TravelAgency struct {
  Id      ActorId
  Address string
  Catalog QueryableCollection[*Journey]
  
  MessageSender MessageSender
}

func (ta *TravelAgency) processTravelBookingRequest(
travelBooking TravelBookingRequest,
) error {
  journey, err := ta.Catalog.Get(travelBooking.JourneyId)
  if err != nil {
    return err
  }
  response := &TravelBookingReply {
    AgencyId:  ta.Id,
    JourneyId: travelBooking.JourneyId,
  }
  if journey.AvailableBookings == 0 {
    response.IsJourneyBooked = false
    response.FailureReason = "Full"
  } else {
    response.IsJourneyBooked = true
    journey.AvailableBookings -= 1
  }
  
  ta.MessageSender.Tell(*response, travelBooking.TravelerId)
  return nil
}
\end{lstlisting}

\texttt{TravelAgency} declares a new collection of travels just by adding a
\texttt{QueryableCollection[*Journey]} field. \sysname execution engine takes
care of injecting the necessary component that is responsible for querying the
collection of travels. When the actor state is loaded, only the \texttt{Id}
and \texttt{Address} fields are fetched from the database: no journey is
loaded. The \texttt{QueryableCollection} component operates lazily, fetching
journeys only when requested. The \texttt{Get} operation fetches the state of
the journey from the database and keeps a cached value of its state.

The state held by each actor is private, so no other actor can modify it. This
property ensures that the cache of items kept by the
\texttt{QueryableCollection} is never stale: every change to its elements is
done by the same actor, so it is not possible to lose any update.
Since the cache always contains the latest version of an item, the
\texttt{Get} method of the \texttt{QueryableCollection} can avoid hitting the
database if the requested item is in the cache.

\fakeparagraph{Find}
The \texttt{Find} API allows to efficiently query a collection for all items
satisfying some property, as exemplified in the following code snippet.

\begin{lstlisting}
type DiscountRequest struct {
  Destination string
  Discount    float64
}

func (ta *TravelAgency) ReceiveMessage(
  message Message
) error {
  // ...
  else if discountRequest, ok := message.(*DiscountRequest); ok {
    return ta.applyDiscount(*discountRequest)
  }
  // ...
}

func (ta *TravelAgency) applyDiscount(
  discountRequest DiscountRequest
) error {
  journeysToUpdate, err := ta.Catalog.Find(
    "Destination", discountRequest.Destination
  )
  if err != nil {
    return err
  }

  for _, journey := range journeysToUpdate {
    journey.Cost -= journey.Cost * discountRequest.Discount
  }

  return nil
}
\end{lstlisting}

The \texttt{Journey} \texttt{QueryableItem} exports attribute
\texttt{Destination} through the \texttt{GetQueryableAttributes()} method, so
it can be used within queries, as exemplified within the invocation of
\texttt{Find}.
The query on the destination is done directly on the database using a specific
index for that attribute, so that the database can immediately locate the
interested items.
As for the \texttt{Get} method, all retrieved items reflect their latest
version and the actor code can directly modify the state of the items without
worrying about state management.

\subsection{Actor spawner}

Actors can spawn other actors as part of their processing.  To do so, they
need to declare as part of their state a special component: the
\texttt{ActorSpawner}, which exposes a \texttt{Spawn} method, defined as
follows.

\begin{lstlisting}
Spawn(newActor Actor, partitionName string, instanceId string) (ActorId, error)
\end{lstlisting}

The \texttt{Spawn} method requires a new instance of the type \texttt{Actor},
the name of the partition, and the instance Id to assign to the new actor and
returns the \texttt{ActorId} of the newly created actor.
The following code snippet shows how to extend the \texttt{TravelAgency} actor
so that it can spawn a new \texttt{TravelAgent} actor that might be used by a
traveler for further interactions after a booking has been made.

\begin{lstlisting}
type TravelAgency struct {
  Id      ActorId
  Address string
  Catalog QueryableCollection[*Journey]
  
  MessageSender MessageSender
  ActorSpawner  ActorSpawner
}

func (ta *TravelAgency) processTravelBookingRequest(
travelBooking TravelBookingRequest
) error {
  journey, err := ta.Catalog.Get(travelBooking.JourneyId)
  if err != nil {
    return err
  }
  response := &TravelBookingReply{
    AgencyId: ta.Id,
    JourneyId: travelBooking.JourneyId,
  }
  if journey.AvailableBookings == 0 {
    response.IsJourneyBooked = false
    response.FailureReason = "Full"
  } else {
    response.IsJourneyBooked = true
    travelAgentId, err := ta.ActorSpawner.Spawn(&TravelAgent{},
      ta.Id.PhyPartitionId.PartitionName, uuid.NewString())
    if err != nil {
      return err
    }
    response.TravelAgentId = travelAgentId
    journey.AvailableBookings -= 1
  }

  ta.MessageSender.Tell(response, travelBooking.TravelerId)
  return nil
}
\end{lstlisting}

Method \texttt{processTravelBookingRequest} spawns a new \texttt{TravelAgent}
if the journey has been correctly created, and includes the identifier of the
newly created actor in the response.
The \texttt{Traveler} actor can store the newly created identifier and use it
to communicate with the agent in case of need.
An important property of \texttt{ActorId} is that once an identifier is
assigned to an actor, it will never change, and it is the only piece of
information needed to communicate with the actor itself.

\section{System Overview}
\label{sec:model}

This section explores the high-level system design of \sysname, presenting its
main architectural components, the data model it adopts, and the lifecycle of
actors.

\subsection{High-level view of the system}

\begin{figure}[H]
\centering
\includegraphics[width=\linewidth]{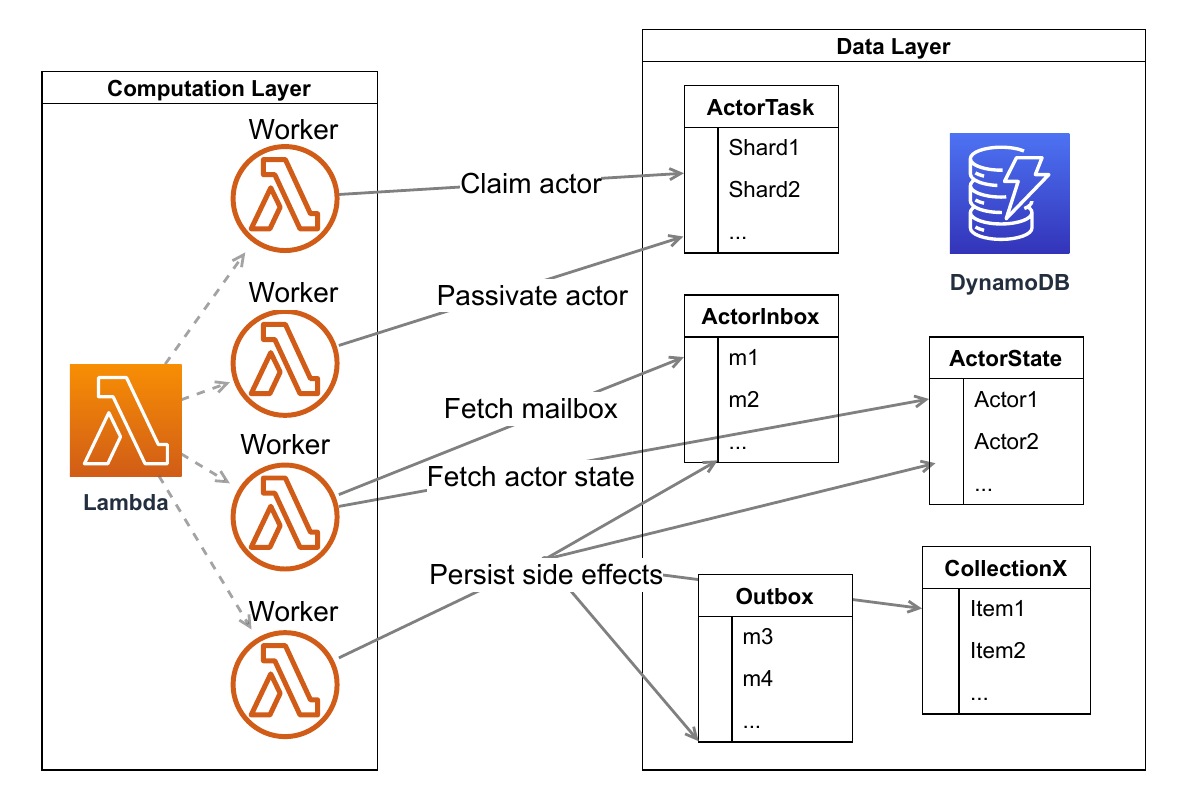}
\caption{Overview of \sysname.}
\label{fig:l-systemOverview}
\end{figure}

\fig{l-systemOverview} presents an overview of \sysname. At a high level,
\sysname consists of a \emph{computation layer} and a \emph{data layer}.

The computation layer consists of \emph{workers}, which are implemented as
instances of Worker Lambda functions.
Workers are responsible for handling the lifecycle of actors, as well as
providing them with all the functionalities they expect from the execution
environment (\texttt{MessageSender}, \texttt{QueryableCollection}, \dots).
Each worker looks for actors that need to be executed, loads them, and then
starts executing their logic.

Workers are executed in Lambda functions, and are designed to run continuously
on invocation. This is different from the most common way to use FaaS, in
which each invocation handles one request and then stops. Workers continue
processing requests until they are terminated. They are tolerant to sudden
shutdowns thanks to our implementation of exactly-once consistency. This
allows external systems to dynamically spawn and stop workers depending on
load conditions without worrying about correctness. 

The data layer is implemented with DynamoDB. It stores the state of actors as
well as inbox and outbox data structures to temporarily store the messages
exchanged between actors.
Workers interact with the data layer during their execution to perform the
following actions:
\begin{inparaenum}[(i)]
\item claiming actors for processing;
\item fetching the state of the actors and their mailboxes;
\item fetching the state of associated \texttt{QueryableCollections};
\item persisting side effects;
\item passivating actors.
\end{inparaenum}

The synchronization between workers is mediated by the data layer: workers
compete when they claim actors for processing, but once an actor has been
claimed, there is no further need for synchronization because an actor can be
processed by at most one worker, and different actors are independent of each
other.
The design of \sysname ensures that workers perform writes and reads on
different partitions of DynamoDB, making the system scalable regarding the
number of workers.

\subsection{Actors organization}

DynamoDB exploits partitioning to improve performance and to scale.
\sysname tries to minimize the impedance between the application and the
database models by introducing partitioning at the application layer.
In particular, actors are organized in \emph{partitions} and \emph{shards}.
A partition is a logical group of actors. The choice of partitions is
domain-specific, and it should be made so that actors that often communicate
with each other end up in the same partition.
A shard is a group of actors within the same partition that are handled
together to reduce the management overhead: actors in the same shard get
claimed together and share the same physical inbox.
While the partitions need to be chosen explicitly by the developer, shards are
automatically handled by \sysname and hence are opaque to the developer.

\subsection{Data model}
\label{sec:model:data}

\sysname uses the following DynamoDB tables to manage the state of the system:
\texttt{ActorTask}, \texttt{ActorInbox}, \texttt{ActorState}, multiple
\texttt{QueriableCollection}s, \texttt{Outbox}.

Before presenting them in details, let us summarize the data model of
DynamoDB.  DynamoDB tables consist of items having a primary key, which must
be unique, and a set of attributes.  Different items are allowed to have
different attributes.

The primary key may consist of a single \emph{partition key} attribute, or a
pair of \emph{partition key}s and \emph{sort key} attributes. In both cases,
the partition key is used to distribute items across multiple hosts: items
with the same partition key are guaranteed to be stored on the same host. In
presence of a sort key, items with the same partition key are stored ordered
by sort key.

\fakeparagraph{ActorTask}
The ActorTask table orchestrates the processing of actor shards. Each item
represents a shard of actors, and uses the \texttt{shard\_id} as primary key.
Its attributes include: \texttt{worker\_id}, which identifies the worker that
claimed the shard, \texttt{insertion\_time}, to prioritize older tasks, and
\texttt{is\_sealed}, a boolean flag used during shard passivation.
An additional \texttt{ActorTaskByWorker} index enables efficient retrieval of
tasks claimed by specific workers, facilitating task acquisition and recovery
after crashes.




\fakeparagraph{ActorInbox}
Message passing between actors is managed through the ActorInbox table, which
stores incoming messages for each shard.  It uses the \texttt{shard\_id} as
its partition key and \texttt{timestamp} as its sort key.
By collapsing mailboxes of actors within the same shard, the system can
retrieve messages for all actors part of a shard with a single query,
significantly reducing the number of database operations.
Each item in this table includes the message \texttt{type},
\texttt{sender\_id}, \texttt{receiver\_id}, and the serialized content of the
message.
The \texttt{timestamp} attributes ensures message ordering, preserving FIFO
guarantees within each actor's communication.




\fakeparagraph{ActorState}
The \texttt{ActorState} table manages the persistency of actors' state.
It uses the \texttt{actor\_id} as its primary key, and it stores the actor
\texttt{type} and the serialized \texttt{current\_state} of each actor,
reflecting the latest committed version of the actor's state.
This structure allows for quick retrieval and updates of actor states during
processing cycles.





\fakeparagraph{QueryableCollection tables}
To support efficient querying of actor-held collections, \sysname implements
\texttt{QueryableCollection} tables. These tables use \texttt{collection\_id}
(a combination of ActorId and field name) as their partition key and
\texttt{item\_id} as their sort key.
They store the state of \texttt{QueryableItem}s and include indexes on
attributes specified by the \texttt{GetQueryableAttributes()} method, enabling
fast lookups over the attributes indicated by the developers.


\fakeparagraph{Outbox}
The \texttt{Outbox} table manages the interaction with external clients. When
submitting requests, clients are responsible for labeling them with a unique
\texttt{correlation\_id}.
The \texttt{Outbox} table stores responses to clients, using the
\texttt{correlation\_id} as its primary key. Its attributes include the
response \texttt{type}, serialized \texttt{content}, \texttt{sender\_id}, and
\texttt{timestamp}.
This table acts as an intermediate buffer, allowing actors to append results
that are sent back to clients asynchronously.


\fakeparagraph{Discussion}
The data model design discussed above optimizes the scalability and
performance of the distributed actor system, balancing the need for efficient
state management, message passing, and query operations.
By leveraging DynamoDB's partitioning and indexing capabilities, \sysname
achieves a robust and flexible architecture capable of handling complex actor
interactions and state management at scale. The design choices, such as
shard-based inboxes and queryable collections, demonstrate a focus on
minimizing database operations while maintaining system consistency and
responsiveness.

\subsection{Actors lifecycle}

\begin{figure}[htbp]
  \centering
  \includegraphics[width=\linewidth]{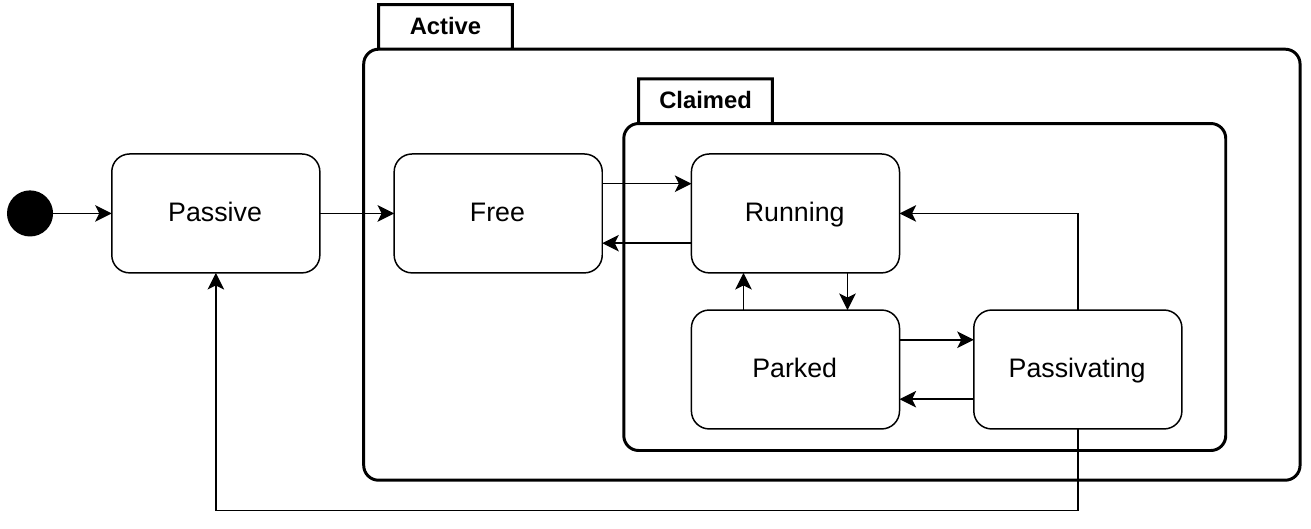}
  \caption{State diagram of a shard}
  \label{fig:l-shardStateDiagram}
\end{figure}

We now present the lifecycle of actors in \sysname. To reduce the overhead of
lifecycle management, actors are organized into shards: all actors within a
shard are treated as an atomic unit and undergo the same states, as
illustrated in \fig{l-shardStateDiagram}.

Shards may be in two macro-states: \emph{active} and \emph{passive}.
Active shards are either being actively processed by a worker (denoted
\emph{claimed}), or they are waiting for a worker to claim them and start the
execution (denoted \emph{free}).
Claimed shards can be in one of three states: \emph{running}, \emph{parked},
or \emph{passivating}. Running shards are actively processing messages, parked
shards have empty inboxes and are checked infrequently, and passivating shards
are in the process of being marked as passive. We distinguish between running
and parked shards to optimize resource usage: the worker can claim more shards
than it can actively process and park the excess to reduce the number of shard
assignment operations. 

The transition from one state to another must take care to ensure the liveness
of the application: it is crucial that any shard with messages in its inbox
eventually becomes active, otherwise messages may be lost.
The most critical transition is that from the passivating state to the passive
state. A naive approach to this transition can lead to race conditions and
potential message loss. To address this, the system employs a sealing strategy
during the passivation process. This strategy ensures that concurrent message
additions to a shard's inbox do not result in the shard being passive with
unprocessed messages.
As the first step of passivation, the shard is atomically marked as sealed
with a flag, from this point onward, any other worker that may try to add a
new message to the inbox of the shard during the passivation procedure will
know that it has to schedule a delayed activation of the shard. In this way,
the passivation procedure can complete, and the message will not be lost as
the sender will ensure to wake up the shard after it has gone passive. 

\section{System implementation}
\label{sec:impl}

We now discuss the implementation details of the \sysname execution
environment.
\s{impl:arch} presents the internal architecture of a worker,
\s{impl:exec-model} describes its execution model, and \s{impl:eos} explains
how \sysname provides exactly-once processing consistency system-wide.

\subsection{Architecture of workers}
\label{sec:impl:arch}

\begin{figure}[hptb]
  \centering
  \includegraphics[width=\linewidth]{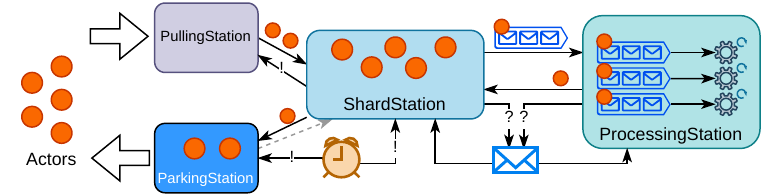}
  \caption{Internal architecture of a worker.}
  \label{fig:l-SystemDiagram}
\end{figure}

Workers are the core components that execute the functionalities of \sysname.
Each worker manages the lifecycle of multiple actors at the same time. To best
utilize available resources, we choose a concurrent, parallel execution model.
Our design originates from a careful analysis of the typical workload of workers:
as workers need to frequently interact with the data layer for communication
and synchronization, they are likely to spend a significant amount of time
waiting for responses from the data layer.
To amortize the time spent waiting for the data layer, we organized the worker
functionalities into concurrent units (called \emph{stations}) that interleave
in computation and yield resources while waiting (see \fig{l-SystemDiagram}).
Each station is responsible for a set of tasks in the lifecycle of shards
(\fig{l-shardStateDiagram}).

The \emph{ShardStation} is central in the architecture and manages the core
logic of the worker. It asks the \emph{PullingStation} for new actors,
it sends them to the \emph{ProcessingStation} for execution, and queues
inactive ones to the \emph{ParkingStation} for passivation.
Each station runs in a goroutine and may delegate sub-tasks to separate
goroutines.
Stations communicate exclusively using Go channels~\cite{channelDefinition}.

\subsection{Execution model}
\label{sec:impl:exec-model}

The ShardStation runs the main control loop that tries to keep a configured
amount of work active on the worker. If it has lower load than it can handle
it sends a request to the PullingStation to fetch new shards for processing
(if any is available).
The PullingStation looks for actors that need processing but are not assigned
to any active worker.
When it finds some, it assigns their shard to the ShardStation. The
ShardStation then polls the database for messages received by the actors that
are part of the assigned shard.

When messages arrive, the ShardStation queues the corresponding actors to be
executed by the ProcessingStation, which runs the actor specific logic by
processing the messages, updating the actor state, and generating output
messages.
All these side effects are written in the data layer with an atomic
transaction. If this transaction fails, no change is applied and the
ProcessingStation rolls back the actor state to what it was before the
execution without consuming the message.

When there are no more messages to be processed for a given actor, the
ProcessingStation notifies the ShardStation.
If a shard has no pending messages for any of its actors, the ShardStation can
decide to park the shard by sending it to the ParkingStation. Here it will
wait, polling for any message that would wake up any of its actors, for a
configured time.  After a time threshold, the ParkingStation starts a
passivation procedure for the shard.
This procedure involves the proper disposal of unused computing resources,
while guaranteeing liveness, meaning that actors with pending messages should
not be left in the passive state.
A shard that has no active actors is marked as \emph{sealed} with an atomic
database operation.  In this state, any other worker adding messages to the
inbox of an actor in a sealed shard will know that it will need to wake the
actor after the passivation process is complete.

A worker's ProcessingStation can be configured to have a variable amount of
processing slots (that are mapped to Go's gorutines). This allows the
processing station to work as a configurable thread pool that can optimize the
utilization of the available resources.

Another key feature that optimizes the utilization of available resources is
the possibility of releasing a shard. When a worker is overloaded and the
queuing time for messages becomes too high, it can release some of its shards.
The released shards will then be acquired by other workers that have available
processing resources, spreading the load and allowing for dynamic scalability
of the system based on demand.

\subsection{Exactly-once consistency}
\label{sec:impl:eos}

\sysname provides exactly-once consistency for messages and side effects: even
in the presence of failures, the system generates the same output and changes
the internal state of actors as if each input request was processed once and
only once with no failures.

At a high level, \sysname provides exactly-once consistency system-wide by
ensuring that each actor processes messages and produces side effects exactly
once.
Specifically, upon processing an input message $m$, an actor can produce side
effects by:
\begin{inparaenum}[(1)]
\item consuming the message $m$ from the actor's inbox; 
\item sending messages to other actors or to the output table;
\item spawning other actors;
\item updating their internal state, including queryable collections.
\end{inparaenum}

\sysname guarantees that the above side effects are executed exactly once by
grouping them together in a single atomic transaction that is committed at the
end of the processing of $m$.  This way either all changes are applied, or no
change is visible, and we can restart the processing without producing
duplicate effects.
To do so, if during the processing of message $m$, an actor wants to send a
set of other messages $m_1 \dots m_n$, we do not send them immediately, but we
put them in a temporary buffer, moving them to the inbox of recipients actors
within the transaction executed at the end of the input message processing.


Crucial to ensure exactly-once consistency is liveness, meaning that any
pending message in the inbox of an actor is eventually consumed and processed.
\sysname guarantees liveness by marking all shards containing actors that are
recipient of a message as active (see \fig{l-shardStateDiagram}). This
information is stored in the ActorTask table within the data layer (see
\s{model:data}).
The ActorTask table maps the assignment of shards to physical workers: if a
shard is in need of processing, it will have an entry in the table, either
with a null worker assignment (signaling that it should be claimed by a
worker) or with the Id of the worker that has claimed it.
\sysname includes the update to the ActorTask table within the transaction
executed at the end of message processing, thus ensuring that all recipients
of outgoing messages are considered for execution.

To avoid duplication of work, all the modifications to the ActorTask table are
done using atomic conditional operations. These operations check a condition
and atomically apply the result only if it was verified. They can be used to
avoid the conflicts when acquiring shards: if two workers were to try and
claim the same shard with no currently assigned workers, only one of them
would obtain the shard, since the conditional check is atomic.

Concerning the spawning of actors, there are two operations that need to be
performed:
\begin{inparaenum}[(1)]
\item create an identifier for the actor;
\item create a new shard that the actor will be part of, only if needed.
\end{inparaenum}
\sysname can immediately create a new shard, since shards are not visible to
the user-facing API and creating a new one cannot affect the logic of the
application.  On the other hand, it delays the creation of the actor
identifier and includes it in the same transaction that contains all other
side effects.

Concerning queryable collections, any write operation to the collection must
follow the same constraints as other side effects. \sysname uses an
actor-local cache of any element an actor fetches from the
collection. Whenever the actor modifies an element, the cache entry
corresponding to that element will be marked as dirty with a flag. At the end
of message processing, we include the update of all dirty cache items in the
atomic transaction, preserving the exactly-once consistency.


To implement these features without requiring effort from the developer we
make use of the reflection capabilities of Go. We define special types, called
\emph{feature types}, that allow actors to use the features. If an actor needs
one of the features we presented in this section, it just needs to include an
attribute with the corresponding feature type in its definition.
When the actor is loaded, \sysname uses reflection to detect these types.
Every feature type is first initialized with its default value, then its
initialization function is executed. These types can interact with the
execution environment and the database to gather all the needed information
and execute operations transparently. The actor code can just assume that the
feature types will be automatically initialized and configured by the system,
hiding all operations required to preserve the exactly-once consistency.

\section{Evaluation}
\label{sec:eval}

We evaluate \sysname to assess whether it fulfills the goals we wanted to
achieve. Specifically, we aim to answer the following research questions:

\begin{itemize}
\item[Q1.] Does \sysname simplify the development of Web applications with
  respect to classical serverless Web development?
\item[Q2.] Does \sysname add significant runtime overhead with respect to
  classical serverless Web solutions?
\item[Q3.] Does \sysname scale well when new workers are added to the system?
\end{itemize}

In this context, we refer to classical serverless development as the
development of Web applications using FaaS and serverless databases manually
coding their interactions.
To answer these questions, we developed two application scenarios. Each
scenario has been implemented both using \sysname and a classical serverless
development methodology (which we will refer to as \emph{baseline} going
forward).

Notice that the baseline implementation provides the typical guarantees of the
traditional serverless model. Specifically, it does not offer the same
exactly-once consistency guarantees as \sysname: in the presence of failures,
requests may be lost or processed more than once, possibly leading to
inconsistent outputs.

To assess Q1, the metrics used are the total number of lines of code and the
percentage of lines of domain logic code with respect to the total lines of
code. To assess Q2, the metrics used are request throughput and latency. To
assess Q3, throughput has been measured varying the number of workers used and
maintaining constant the workload.

\subsection{Experiments setup}
\label{sec:eval:setup}

\fakeparagraph{Application scenarios}
To test \sysname, we selected two application scenarios to include both simple
tasks, where each request is handled by a single actor, and more complex ones,
where each request involves the interaction of different actors. Moreover,
both scenarios include cases in which multiple requests access the same state,
to stress the problem of concurrency control.

The first scenario is a \emph{banking} application, where users can execute
financial transactions between two accounts.
In \sysname, we implemented banks as actors (of type \texttt{Bank}) and
accounts as \texttt{QueryableItem}s of the bank.
Banks receive transaction requests from clients and execute them. The state of
each account must be protected against multiple transactions that involve it.

The second scenario is a \emph{hotel reservation} application, where users can
book rooms asking for a specific room type and booking interval.
In \sysname, we defined two types of actors: \texttt{User} and \texttt{Hotel}.
Users create booking requests and send them to the hotel actors. Hotel actors
check availability for the requested interval and the room type, generate a
reservation, and send the reservation back to the users. After inspecting the
reservation, users terminate the interaction. The availability of each hotel
must be protected against multiple booking requests for the same hotel.


\fakeparagraph{Evaluation environment}
We evaluated the solutions in the AWS ecosystem, using AWS Lambda for FaaS and
DynamoDB as serverless database.
We configured DynamoDB in On-Demand mode, which offers a pay-per-request
pricing model, leading to predictable costs that are directly proportional to
the number of operations performed.
A downside of the On-Demand mode is that, while AWS states that DynamoDB can
scale up to sustain any workload, it can take as long as 30 minutes of high
activity before the scaling process completes.
To limit costs and ease reproducibility, we ran each experiment for
approximately 2 minutes, so the maximum throughput available was the default
one provisioned by DynamoDB, around 4000 single write operations per second
and 12000 read operations per
second.\footnote{\url{https://docs.aws.amazon.com/amazondynamodb/latest/developerguide/on-demand-capacity-mode.html}}

In our evaluation, we spawn a static number of workers. Workers are fault
tolerant and automatically rebalance shards to spread the load. Because of
this, future work could easily add an external dynamic scheduler to start and
stops workers depending on demand and resource utilization.

\fakeparagraph{Evaluation metrics}
We used the following metrics to measure the performance of \sysname and the
simplicity of its programming model.

\smallskip \noindent
\emph{Lines of code.}
Lines of code can vary between two functionally equivalent implementations of
the same problem, so this metric must be taken into consideration with care.
To mitigate possible bias between the \sysname implementation and the
baseline, the domain code of the two solutions has been written as similarly
as possible.
The main differences come from data access patterns and concurrency control,
so we split the lines of the implementations into two categories: domain code
and infrastructure interaction.
We measure the percentage of domain code over total code, giving an estimation
of the effort spent writing code for the problem at hand with respect to
handling infrastructure interaction.

\smallskip \noindent
\emph{Throughput.}
To evaluate throughput, we feed a fixed number of requests to the system and
measure the time required to process all of them and the time each request has
been completed. Using this data we can extrapolate the average throughput and
its trend over time.
For the banking scenario, the system has been fed with 60k requests spanning
30k bank accounts.
For the hotel reservation scenario, the system has been fed with 10k requests
involving 200 users and 100 hotels resulting in a high resource contention.

\smallskip \noindent
\emph{Latency.}
To evaluate latency, we feed requests at a fixed rate (below the maximum input
rate the system can sustain), and we measure the time required to fulfill each
request. We measure the latency as the interval between the point in time when
the request starts to be processed and the point in time when the
corresponding output is produced.
For both scenarios, we consider an input rate of 5 requests per second, and we
run each experiment for 2 minutes.

An important parameter of the \sysname implementation is the polling interval.
\sysname will check for new messages by polling at a fixed time interval.
To evaluate the effect of this parameter on the final latency, the \sysname
implementation has been run with the same workload with different values for
polling interval: 100, 500 and 1000 ms.
We make the comparison fair by adding half of the polling interval to the
reported latency, this accounts for the mean waiting time for the input
message to be detected by the system.

\smallskip \noindent
\emph{Scalability.}
To study scalability, we measured the throughput of \sysname under the same
workload, while changing the number of workers (strong scalability).
We consider 1, 2, 4, and 8 workers.

\subsection{Results}
\label{sec:eval:results}

We now present the results we measured for the two applications scenarios,
considering the evaluation metrics discussed in \s{eval:setup}.

\fakeparagraph{Banking scenario}

\smallskip \noindent
\emph{Lines of code.}
\tab{l-bankingLoC} shows the total lines of code and the percentage of domain
code for \sysname and the baseline.
\sysname reduces the total lines of code by a factor of two with respect to
the baseline.  Most significantly, the entire application only contains 17
lines of infrastructure code, meaning that over 80\% of the lines are used to
implement domain logic.  In comparison, the baseline requires 125 lines of
infrastructure code, which account for almost 65\% of the total codebase for
the banking application.

\begin{table}[hptb]
  \scriptsize
  \begin{center}
    \begin{tabular}{l|c|c|c}
      \textbf{System} & \textbf{Infrastructure} & \textbf{Total} & \textbf{Domain code} \\
      \hline
      \sysname        & 17                      & 94             & 81.9\% \\
      Baseline        & 125                     & 195            & 35.9\% \\
    \end{tabular}
    \caption{Lines of code (infrastructure and total) for the banking
      scenario.}
    \label{tab:l-bankingLoC}
  \end{center}
\end{table}

This gap is due to the ability of \sysname to handle state management and
querying automatically and transparently. Moreover, the actor model inherently
avoids low-level race conditions on the bank accounts, so there is no need to
explicitly lock the accounts involved in a transaction to guarantee
safety. Instead, part of the code of the baseline is uniquely dedicated to
setting up the queries to DynamoDB and extracting results or ensuring the
correctness of the transaction by locking the accounts involved.
This analysis is confirmed by the breakdown of the infrastructure lines of
code for both implementations, as shown in \tab{l-bankingInfrastructureCode}.

\begin{table}[hptb]
  \scriptsize
  \begin{center}
    \begin{tabular}{l|l|c}
      \textbf{System} & \textbf{Functionality} & \textbf{LoC} \\
      \hline
      \multirow{3}{*}{\sysname} & Declaration of infrastructure fields and methods & 13 \\
                                & QueryableCollection API & 2 \\
                                & Communication & 2 \\
      \hline
      \multirow{2}{*}{Baseline} & DynamoDB API & 97 \\
                                & Locking and retrying & 28 \\
    \end{tabular}
    \caption{Infrastructure code breakdown for the banking scenario.}
    \label{tab:l-bankingInfrastructureCode}
  \end{center}
\end{table}

\smallskip \noindent
\emph{Throughput and scalability.}

\begin{figure}[hptb]
  \centering
  \includegraphics[width=0.9\linewidth]{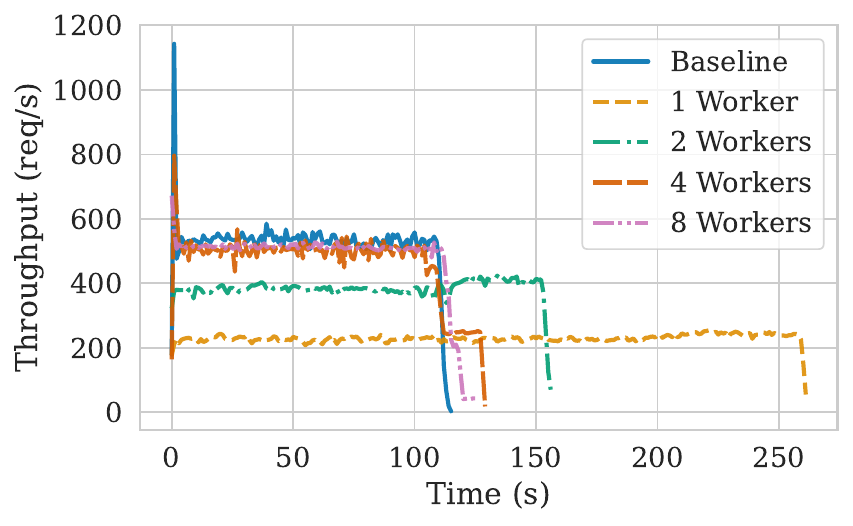}
  \caption{Throughput for the banking scenario.}
  \label{fig:l-bankingThroughput}
\end{figure}

\fig{l-bankingThroughput} shows the throughput we measured in the banking
scenario for \sysname and for the baseline.  For \sysname, we use 4 different
configurations, varying the number of workers used.
As \fig{l-bankingThroughput} shows, the throughput of \sysname grows linearly
with respect to the number of workers, except from 4 workers onward. With 4 or
more workers, the system saturates the throughput of the database and DynamoDB
starts to throttle requests. At that point, the database becomes the
bottleneck and further workers do not improve throughput as shown by the
similar traces of the 4 and 8 workers setup.
DynamoDB becomes the bottleneck for the baseline as well, and the saturated
throughput is around 400 and 600 transactions per second.

\smallskip \noindent
\emph{Latency.}

\begin{figure}[hptb]
  \centering
  \includegraphics[width=0.9\linewidth]{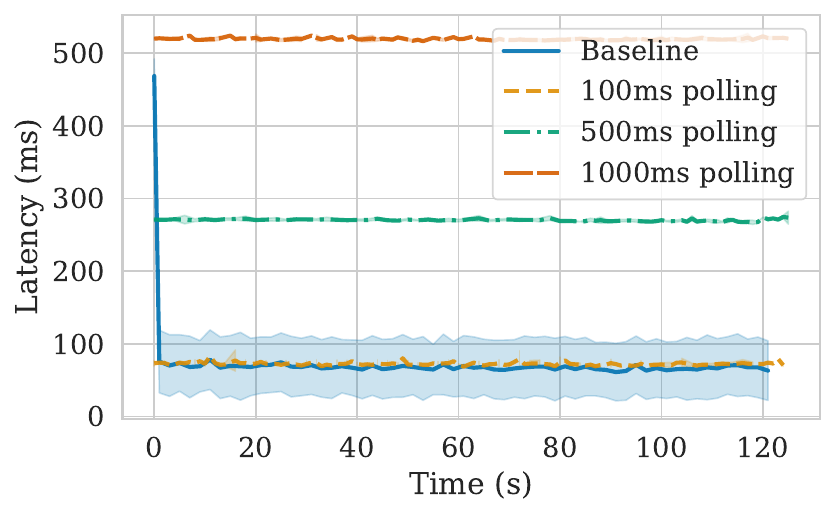}
  \caption{Latency for the banking scenario.}
  \label{fig:l-bankingLatency}
\end{figure}

\fig{l-bankingLatency} shows the latency we measured in the banking scenario
for \sysname and for the baseline.
For \sysname, two factors contribute to the total latency:
\begin{inparaenum}[(i)]
\item the time the request waits before a worker starts to actively process
  it, and
\item the actual processing time.
\end{inparaenum}
The first time depends on the polling time, which is a configuration parameter
developers can set to trade response time for costs. Frequent polling reduces
the time requests wait before being processing, at the cost of more
interactions with the database. On average, messages wait for half of the
polling time.
Processing time is lower for \sysname than for the baseline (22.3ms vs 74ms,
on average): this can be attributed to the lack of low-level race conditions
in the actor model.
As a result, the latency of \sysname is dominated by waiting time in all the
configurations we tested (with polling time ranging between 100ms and
1000ms). With a polling time of 100ms, the latency of \sysname and the
baseline are almost identical.

\fakeparagraph{Hotel reservation scenario}

\smallskip \noindent
\emph{Lines of code.}

\tab{l-hotelLoC} shows the total number of lines of code and the percentage of
infrastructure code for \sysname and the baseline in the hotel reservation
scenario. The increased size of the codebase with respect to the banking
scenario indicate a higher complexity.
Also in this case, \sysname shows a tangible improvement with respect to the
baseline both in terms of total number of lines of code (233 vs 356) and, most
importantly, in terms of infrastructure code (33 vs 169).

\begin{table}[hptb]
  \scriptsize
  \begin{center}
    \begin{tabular}{l|c|c|c}
      \textbf{Implementation} & \textbf{Infrastructure} & \textbf{Total} & \textbf{Domain code} \\
      \hline
      \sysname                & 33                      & 233            & 85.8\%               \\
      Baseline                & 169                     & 356            & 52.5\%               \\
    \end{tabular}
    \caption{Lines of code (infrastructure and total) for the hotel
      reservation scenario.}
    \label{tab:l-hotelLoC}
  \end{center}
\end{table}

Once again the reduction can be attributed to the programming model of
\sysname, which abstracts away most of the concerns related to the
infrastructure.
\tab{l-hotelInfrastructureCode} shows the breakdown of the infrastructure
lines of code. For the baseline, most of the lines (157 lines) involve
interactions with the database, and a small part (12 lines) involve locking
and retrying methods to implement concurrency control.
For \sysname, most of the lines (28 lines) involve declaring fields and
methods to handle the state of actors and the communication between actors.  A
small part is used to encode the communication between actors (4 lines) and to
access a collection (1 line).

\begin{table}[hptb]
  \scriptsize
  \begin{center}
    \begin{tabular}{l|l|c}
      \textbf{Implementation} & \textbf{Functionality} & \textbf{LoC} \\
      \hline
      \multirow{2}{*}{\sysname} & Declaration of infrastructure fields and
      methods                   & 28 \\
                                & QueryableCollection API & 1 \\
                                & Communication & 4 \\
      \hline
      \multirow{2}{*}{Baseline} & DynamoDB API calls & 157 \\
                                & Locking and retrying mechanism & 12 \\
    \end{tabular}
    \caption{Infrastructure code breakdown for the hotel reservation scenario.}
    \label{tab:l-hotelInfrastructureCode}
  \end{center}
\end{table}

\smallskip \noindent
\emph{Throughput and scalability}

\begin{figure}[hptb]
  \centering
  \includegraphics[width=0.9\linewidth]{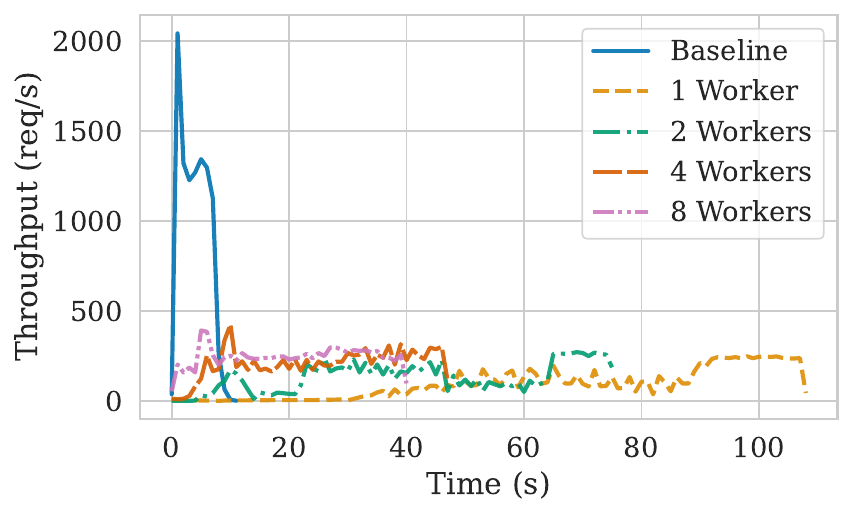}
  \caption{Throughput for the hotel reservation scenario.}
  \label{fig:l-hotelThroughput}
\end{figure}

\fig{l-hotelThroughput} shows the throughput over time for \sysname and the
baseline in the hotel reservation scenario, and \tab{l-avgThroughputs}
summarizes the average throughput for the two systems.  For \sysname, we
considered 4 configurations with different number of workers.

\begin{table}[hptb]
  \scriptsize
  \begin{center}
    \begin{tabular}{c|c}
      \textbf{Configuration} & \textbf{Average throughput} \\
                             & req/s                       \\
      \hline
      1 Worker               & 92.6                        \\
      2 Workers              & 133.3                       \\
      4 Workers              & 212.7                       \\
      8 Workers              & 250.0                       \\
      Baseline               & 909.1                       \\
    \end{tabular}
    \caption{Average throughput for the hotel reservation scenario.}
    \label{tab:l-avgThroughputs}
  \end{center}
\end{table}

Again, in this scenario we see the system scaling and increasing the
throughput as the number of workers increases. However, when comparing with
the baseline implementation, we see lower measured throughput. Remember,
however, that \sysname provides higher guarantees with respect to the
baseline, which may not preserve exactly-once consistency in the presence of
failures.

When implementing this specific scenario in \sysname, each input request
involves the execution of multiple actors that interact by exchanging
messages. To preserve exactly-once consistency (see \s{impl:eos}), each
step of processing requires executing an atomic transaction. As a consequence,
the data layer becomes a bottleneck and prevents linear scaling of throughput
with the number of workers.
This overhead may thus decrease by configuring DynamoDB with a higher
capacity.

\smallskip \noindent
\emph{Latency.}

\begin{figure}[hptb]
  \centering
  \includegraphics[width=0.9\linewidth]{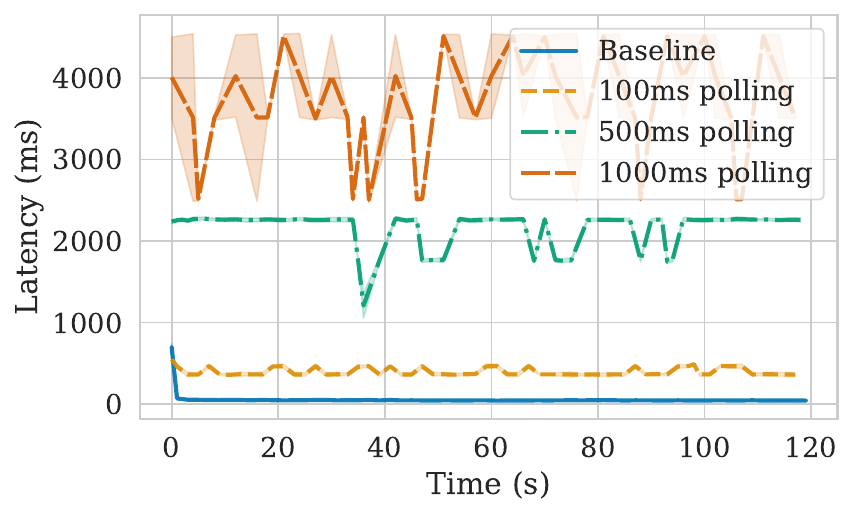}
  \caption{Latency for the hotel reservation scenario.}
  \label{fig:l-hotelLatency}
\end{figure}

\fig{l-hotelLatency} shows the latency measured for \sysname and the baseline
in the hotel reservation scenario.
To understand the results, let us summarize the interactions between actors in
the hotel reservation scenario.
When a client issues a booking request $r_1$, this is placed in the inbox of
the user actor $u$.
When actor $u$ processes the request, it sends a booking message $r_2$ to the
hotel actor $h$.
After a certain amount of time $\Delta t_{1}$, actor $h$ reads $r_2$ from its
inbox and places the response $r_3$ into the inbox of $u$.
After a certain amount of time $\Delta t_{2}$, actor $u$ reads from its inbox
$r_3$, processes it and terminates the request.
The latency of said request is:

\begin{equation*}
latency = p_1 + \Delta t_{1} + p_2 + \Delta t_{2} + p_3
\end{equation*}

\noindent
where $p_1, p_2$, and $p_3$ are the processing time for each message.  We
measured the processing times to be around 20ms, so the main contributors to
the latency are $\Delta t_{1}$ and $\Delta t_{2}$.
These time intervals depend on the polling frequency of the workers: the
smaller the polling interval, the faster a worker (and consequently an actor)
can read a new message from its inbox.
As \fig{l-hotelLatency} shows, smaller polling intervals bring \sysname closer
to the baseline solution. However, in this scenario, the presence of multiple
interactions between actors introduces a latency that is higher than the
baseline when considering the polling intervals we adopted in our experiments.

\subsection{Discussion}
\label{sec:eval:discussion}

Based on the results, it is possible to address the initial research
questions.

\begin{itemize}
\item[Q1.] \sysname significantly simplifies the development of Web solutions
  with respect to classic serverless Web development. It does so by offering
  developers a model suitable for building concurrent applications and a set
  of features to transparently handle state management and querying.
\item[Q2.] \sysname performance heavily depend on the polling interval, which
  is a parameter that developers can use to trade performance for costs. With
  smaller polling intervals, it can provide comparable latency with respect to
  classic serverless development.  However, longer polling time may introduce
  non-negligible overhead in terms of latency in complex scenarios.
  It must be noted, however, that the baseline implementation we compare to
  does not provide the same exactly-once consistency guarantees \sysname
  offers by default.
  %
\item[Q3.] \sysname scales almost linearly when varying the number of workers
  between 1 and 4, unless limited by the data layer.
\end{itemize}

These results indicate that \sysname provides a good programming abstraction
for Web applications.  The performance overhead may be reduced by provisioning
more resources to the data layer, and it can be controlled through the polling
interval.


\section{Related work}
\label{sec:related}

The growing complexity and volume of Web applications and the availability of
new execution environments led researchers and companies to investigate ways
to simplify the development process.
One of the many directions of research is related to the serverless ecosystem
and how to offer abstractions over serverless stateful functions (SSF), which
are serverless functions that can be used as stateful components by accessing
a database.
Other lines of research related to our work focus more specifically on actor
systems.
This section explores some works in these fields.

\subsection{Serverless and stateful functions}

The following projects aim to offer guarantees or programming abstractions
over stateless serverless functions.
\tab{l-relatedWorksComparison} shows a comparison of the solutions presented
in this section.
Each solution has been evaluated along three main dimensions.

\begin{itemize}
\item State management: does the system support integrated state management,
  allowing components of the system to persist and manage their state?
\item Concurrency: does the system allow components to run concurrently
  without compromising consistency guarantees?
\item Transparency: does the system hide the mechanisms through which state
  management and/or concurrency are offered?
\end{itemize}

\begin{table*}[hptb]
  \scriptsize
  \begin{center}
    \begin{tabular}{l|c|c|c}
      & \textbf{State management}  & \textbf{Concurrency} & \textbf{Transparency} \\ 
      \hline
      Nubes            & External                   & Synchronization is needed  & Fully transparent \\
      Oparca           & Mixed (DHT and datastores) & Localized locking          & Fully transparent \\
      Durable Entities & External                   & Serial processing          & Fully transparent \\
      Kalix            & External                   & Serial processing          & Fully transparent \\
      CloudBurst       & External                   & Causal consistency         & Explicit API      \\
      Beldi            & External                   & Synchronization primitives & Explicit API      \\
      Boki             & Internal (storage nodes)   & Synchronization primitives & Explicit API      \\
      Apache StateFun  & Internal                   & Serial processing          & Fully transparent \\
      Faasm            & Internal (two-tier)        & Locking mechanism          & Explicit API      \\
      Apiary           & Internal                   & Transactional guarantees   & Explicit API      \\
      Crucial          & Internal (DSO layer)       & Linearizable objects       & Explicit API      \\
      \sysname         & Mixed (external and cache) & Serial processing          & Fully transparent \\
    \end{tabular}
    \caption{Comparison of programming abstractions for serverless stateful
      functions.}
    \label{tab:l-relatedWorksComparison}
  \end{center}
\end{table*}

Nubes~\cite{nubesArticle} introduces an Object-Oriented Programming (OOP)
abstraction layer for stateful serverless functions, allowing developers to
define types with methods that are transparently executed as cloud
functions. While Nubes employs OOP concepts, our system leverages the actor
model, which inherently handles synchronization issues that would require
explicit management in an OOP paradigm. This choice simplifies concurrent
programming and reduces the likelihood of race conditions.

Oparca~\cite{oparcaPaper} implements the Objects-as-a-Service (OaaS) paradigm,
managing object lifecycles and method invocations through a dedicated Invoker
component. It ensures consistency between structured and unstructured states,
and provides exactly-once guarantees for asynchronous method invocations. Our
system shares Oparca's commitment to consistency and exactly-once semantics
but achieves this within the more flexible actor model framework.

Azure's Durable Entities~\cite{durableEntitiesDocs}, part of the Durable
Functions service, implement actor-like components within the Azure Functions
ecosystem. These entities provide durability of actor state and reliable
messaging. However, they lack the comprehensive query capabilities offered by
our system. Our approach not only ensures state durability but also enables
complex data aggregation and analysis directly on the actor state, a crucial
feature for many distributed applications.

Kalix~\cite{kalixSite} offers high-level abstractions such as entities, views,
actions, and workflows to model business domains, managing state and
concurrency transparently. While Kalix provides a complete platform, our
system gives flexible abstractions that can be adapted to different
technologies and integrate with existing infrastructures.

CloudBurst~\cite{cloudBurstArticle} enhances stateless functions with
efficient state transfer and point-to-point communication capabilities. It
leverages a combination of key-value stores and local caches to achieve low
latency. Our system goes beyond CloudBurst by providing not just efficient
state management but also a complete actor-based programming model with
built-in consistency guarantees and querying capabilities.

Beldi~\cite{beldiArticle} and Boki~\cite{bokiArticle} focus on providing
exactly-once semantics for function invocations. Beldi introduces the concept
of Intent, while Boki improves upon this with a shared log abstraction and
optimized read caches. Our system incorporates similar reliability guarantees
within the actor model framework, offering a more comprehensive solution that
includes not just exactly-once processing but also robust state management and
querying.

Apache Flink Stateful Functions~\cite{flinkDocs} is an environment that
simplifies the development of distributed applications. It offers exactly-once
processing guarantees by pairing a Flink cluster to a FaaS system. On the
contrary, our system does not require an external managed system to enact the
exactly-once semantics: all the key components run within the autoscaling FaaS
environment. Furthermore, Flink StateFun lacks built-in querying capabilities
and the programming facilities provided by \sysname.

Faasm~\cite{faasmArticle} is a FaaS environment that aims to provide efficient
stateful computing. It introduces a lightweight isolation mechanism that
allows different function instances to run on the same host.  Faasm adopts a
two-tier state architecture: local memory enables efficient shared memory
within a single host, and global memory is used to communicate between hosts.
It offers convenient API to implement Distributed Data Objects (DDOs), an
abstraction that hides the complexity of the two-tier state architecture.
The innovation brought by Faasm is orthogonal with respect to \sysname, and we
could exploit its lightweight FaaS environment as a target for our programming
abstraction.

Apiary~\cite{apiaryPaper} co-locates compiled functions with storage nodes to
optimize computation-storage interaction.  It enables defining groups of
functions that are executed with transactional semantics.
Crucial treats serverless functions as cloud threads, allowing developers to
write applications as standard multithreaded programs.
Our approach differs by providing a higher-level actor-based abstraction that
simplifies distributed system development.

\subsection{Actor Systems}

Akka~\cite{akkaDocumentation}, a prominent implementation of the actor model
for JVM environments, offers a modular approach to building distributed
systems. Its approach to state management and querying differs from ours,
offering optional persistence through snapshots or event
sourcing~\cite{eventSourcingPattern}. In contrast, our system provides
always-durable state storage and direct state querying, ensuring strong
consistency without the need for additional modules or configurations.

Orleans~\cite{orleansDocs} introduces the concept of Virtual Actors, which are
always conceptually present and instantiated on-demand. It offers location
transparency similar to our system but lacks built-in query functionalities
and provides weaker message delivery guarantees. Our system enhances the
virtual actor concept with strong consistency guarantees and comprehensive
querying capabilities, addressing key limitations in Orleans' approach.

Our system builds upon these foundations, combining the strengths of
actor-based models with serverless architectures. It offers durable state
management, powerful querying capabilities, fault-tolerance, and concurrency
control, addressing the limitations of existing solutions in a unified
framework.

\section{Conclusions}
\label{sec:conclusions}

With \sysname, we brought the actor programming paradigm to serverless
environments.
\sysname relieves developers from the burden of interacting with external
storage systems, ensuring state consistency and durability, and adopting
concurrency control mechanisms.
\sysname abstracts away these concerns: the actor model prevents data races by
default and the messaging protocol of \sysname makes sure that the system
behaves as if all requests were processed once and only once, even in the
presence of failures.
Additionally, \sysname provides convenient functionalities to easily and
efficiently query state.

The effectiveness of \sysname has been measured in two benchmark scenarios
comparing it to baseline implementations of the same scenarios. The results
showed a significant reduction of coding overhead, with a high percentage of
lines of code used to model domain logic. This confirms the ability of
\sysname to simplify the process of web development.
The benchmarks confirmed the scalability of the system at least up until the
saturation of the available resources.  The performance overhead of the system
can be controlled with a configuration parameter that balances performance and
costs.

Motivated by the effectiveness of our programming model in serverless
environments, we plan to investigate alternative implementation strategies to
further improve performance and make \sysname even more beneficial for
developers: for instance, adopting services that enable a reactive interaction
between actors could factor out the overhead of polling.


\bibliographystyle{IEEEtran}

\begin{thebibliography}{10}
\providecommand{\url}[1]{#1}
\csname url@samestyle\endcsname
\providecommand{\newblock}{\relax}
\providecommand{\bibinfo}[2]{#2}
\providecommand{\BIBentrySTDinterwordspacing}{\spaceskip=0pt\relax}
\providecommand{\BIBentryALTinterwordstretchfactor}{4}
\providecommand{\BIBentryALTinterwordspacing}{\spaceskip=\fontdimen2\font plus
\BIBentryALTinterwordstretchfactor\fontdimen3\font minus
  \fontdimen4\font\relax}
\providecommand{\BIBforeignlanguage}[2]{{%
\expandafter\ifx\csname l@#1\endcsname\relax
\typeout{** WARNING: IEEEtran.bst: No hyphenation pattern has been}%
\typeout{** loaded for the language `#1'. Using the pattern for}%
\typeout{** the default language instead.}%
\else
\language=\csname l@#1\endcsname
\fi
#2}}
\providecommand{\BIBdecl}{\relax}
\BIBdecl

\bibitem{actorModelDefinition}
G.~Agha, \emph{Actors: a model of concurrent computation in distributed
  systems}.\hskip 1em plus 0.5em minus 0.4em\relax Cambridge, MA, USA: MIT
  Press, 1986.

\bibitem{serverlessDefinition}
{Red Hat}, ``What is serverless?'' 2022,
  \url{https://www.redhat.com/en/topics/cloud-native-apps/what-is-serverless}.

\bibitem{faasDefinition}
------, ``What is function-as-a-service (faas)?'' 2020,
  \url{https://www.redhat.com/en/topics/cloud-native-apps/what-is-faas}.

\bibitem{faasImplications}
M.~Shahrad, J.~Balkind, and D.~Wentzlaff, ``Architectural implications of
  function-as-a-service computing,'' in \emph{Proceedings of the International
  Symposium on Microarchitecture}, ser. MICRO '52.\hskip 1em plus 0.5em minus
  0.4em\relax New York, NY, USA: ACM, 2019, p. 1063–1075.

\bibitem{awsLambda}
Amazon, ``Aws lambda,'' 2024, \url{https://aws.amazon.com/lambda/}.

\bibitem{azureFunctions}
Microsoft, ``Azure functions,'' 2024,
  \url{https://azure.microsoft.com/products/functions}.

\bibitem{cloudFunctions}
Google, ``Cloud functions,'' 2024, \url{https://cloud.google.com/functions}.

\bibitem{cloudflareWorkers}
Cloudflare, ``Cloudflare workers,'' 2024, \url{https://workers.cloudflare.com}.

\bibitem{serverlessDatabaseDefinition}
{Amazon Web Services}, ``What is a serverless database?'' 2024,
  \url{https://aws.amazon.com/what-is/serverless-database/}.

\bibitem{awsDynamoDB}
Amazon, ``Aws dynamodb,'' 2024, \url{https://aws.amazon.com/dynamodb/}.

\bibitem{dynamodbLatency}
{Sandip Gangdhar}, ``Understanding amazon dynamodb latency,'' 2023,
  \url{https://aws.amazon.com/blogs/database/understanding-amazon-dynamodb-latency/}.

\bibitem{channelDefinition}
Google, ``Effective go - channels,'' 2009,
  \url{https://go.dev/doc/effective_go#channels}.

\bibitem{nubesArticle}
K.~A. Marek, L.~De~Martini, and A.~Margara, ``Nubes: Object-oriented
  programming for stateful serverless functions,'' in \emph{Proceedings of the
  International Workshop on Serverless Computing}, ser. WoSC '23.\hskip 1em
  plus 0.5em minus 0.4em\relax New York, NY, USA: ACM, 2023, p. 30–35.

\bibitem{oparcaPaper}
\BIBentryALTinterwordspacing
P.~Lertpongrujikorn and M.~A. Salehi, ``Object as a service: Simplifying
  cloud-native development through serverless object abstraction,'' 2024.
  [Online]. Available: \url{https://arxiv.org/abs/2408.04898}
\BIBentrySTDinterwordspacing

\bibitem{durableEntitiesDocs}
Microsoft, ``Entity functions,'' 2023,
  \url{https://learn.microsoft.com/en-us/azure/azure-functions/durable/durable-functions-entities}.

\bibitem{kalixSite}
{Lightbend}, ``{Kalix},'' 2024, \url{https://www.kalix.io/}.

\bibitem{cloudBurstArticle}
\BIBentryALTinterwordspacing
V.~Sreekanti, C.~Wu, X.~C. Lin, J.~Schleier-Smith, J.~E. Gonzalez, J.~M.
  Hellerstein, and A.~Tumanov, ``Cloudburst: stateful functions-as-a-service,''
  \emph{Proc. VLDB Endow.}, vol.~13, no.~12, p. 2438–2452, 2020. [Online].
  Available: \url{https://doi.org/10.14778/3407790.3407836}
\BIBentrySTDinterwordspacing

\bibitem{beldiArticle}
H.~Zhang, A.~Cardoza, P.~B. Chen, S.~Angel, and V.~Liu, ``Fault-tolerant and
  transactional stateful serverless workflows,'' in \emph{USENIX Symposium on
  Operating Systems Design and Implementation}, ser. OSDI '20.\hskip 1em plus
  0.5em minus 0.4em\relax USENIX Association, 2020, pp. 1187--1204.

\bibitem{bokiArticle}
Z.~Jia and E.~Witchel, ``Boki: Towards data consistency and fault tolerance
  with shared logs in stateful serverless computing,'' \emph{ACM Transactions
  on Computer Systems}, 2024.

\bibitem{flinkDocs}
{Apache Software Foundation}, ``{Stateful Functions: A Platform-Independent
  Stateful Serverless Stack},'' 2024,
  \url{https://nightlies.apache.org/flink/flink-statefun-docs-stable/}.

\bibitem{faasmArticle}
S.~Shillaker and P.~Pietzuch, ``Faasm: Lightweight isolation for efficient
  stateful serverless computing,'' in \emph{USENIX Annual Technical
  Conference}, ser. USENIX ATC 20.\hskip 1em plus 0.5em minus 0.4em\relax
  USENIX Association, 2020, pp. 419--433.

\bibitem{apiaryPaper}
\BIBentryALTinterwordspacing
P.~Kraft, Q.~Li, K.~Kaffes, A.~Skiadopoulos, D.~Kumar, D.~Cho, J.~Li,
  R.~Redmond, N.~Weckwerth, B.~Xia, P.~Bailis, M.~Cafarella, G.~Graefe,
  J.~Kepner, C.~Kozyrakis, M.~Stonebraker, L.~Suresh, X.~Yu, and M.~Zaharia,
  ``Apiary: A dbms-integrated transactional function-as-a-service framework,''
  2023. [Online]. Available: \url{https://arxiv.org/abs/2208.13068}
\BIBentrySTDinterwordspacing

\bibitem{akkaDocumentation}
Lightbend, ``Akka documentation,'' 2024, \url{https://akka.io/docs/}.

\bibitem{eventSourcingPattern}
M.~Fowler, ``Event sourcing,'' 2005,
  \url{https://martinfowler.com/eaaDev/EventSourcing.html}.

\bibitem{orleansDocs}
Microsoft, ``{Microsoft Orleans},'' 2024,
  \url{https://learn.microsoft.com/en-us/dotnet/orleans/overview}.

\end{thebibliography}


\end{document}